\documentclass[aps,prl,twocolumn,superscriptaddress,raggedbottom,showpacs,floatfix]{revtex4}
\usepackage{amsmath,amsfonts,amssymb,amsthm}
\usepackage{bm}
\usepackage{color}
\usepackage{graphicx,latexsym}
\usepackage{epstopdf}
\usepackage{appendix}
\usepackage{float}

\begin{document}
\renewcommand{\vec}{\mathbf}
\renewcommand{\Re}{\mathop{\mathrm{Re}}\nolimits}
\renewcommand{\Im}{\mathop{\mathrm{Im}}\nolimits}

\title{Charged skyrmions on the surface of a topological insulator}
\author{Hilary M. Hurst}
\affiliation{Joint Quantum Institute and Condensed Matter Theory Center, Department of Physics, University of Maryland, College Park, Maryland 20742-4111, USA}
 \affiliation{School of Physics, Monash University, Melbourne, Victoria 3800, Australia}
\author{Dmitry K. Efimkin}
\affiliation{Joint Quantum Institute and Condensed Matter Theory Center, Department of Physics, University of Maryland, College Park, Maryland 20742-4111, USA}
 \affiliation{School of Physics, Monash University, Melbourne, Victoria 3800, Australia}
\author{Jiadong Zang}
\affiliation{Institute for Quantum Matter, Department of Physics and Astronomy, Johns Hopkins University, Baltimore, Maryland 21218, USA}
\author{Victor Galitski}
\affiliation{Joint Quantum Institute and Condensed Matter Theory Center, Department of Physics, University of Maryland, College Park, Maryland 20742-4111, USA}
 \affiliation{School of Physics, Monash University, Melbourne, Victoria 3800, Australia}

\begin{abstract}
We consider the interplay between magnetic skyrmions in an insulating thin film and the Dirac surface states of a 3D topological insulator (TI), coupled by proximity effect. The magnetic texture of skyrmions can lead to confinement of Dirac states at the skyrmion radius, where out of plane magnetization vanishes. This confinement can result in charging of the skyrmion texture. The presence of bound states is robust in an external magnetic field, which is needed to stabilize skyrmions. It is expected that for relevant experimental parameters skyrmions will have a few bound states that can be tuned using an external magnetic field. We argue that these charged skyrmions can be manipulated directly by an electric field, with skyrmion mobility proportional to the number of bound states at the skyrmion radius. Coupling skyrmionic thin films to a TI surface can provide a more direct and efficient way of controlling skyrmion motion in insulating materials. This provides a new dimension in the study of skyrmion manipulation.
\end{abstract}
\maketitle
\emph{Introduction.} Topological insulators (TIs) are a new class of matter with a band structure that can be characterized by the topological invariant (See Refs.~\cite{Hasan2010, Qi2011} and references therein). As with usual insulators, TIs have a forbidden band gap in the bulk separating the filled valence and empty conduction band. Contrary to the former, the topology dictates them to posses very unusual edge (2D) or surface (3D) states protected from nonmagnetic disorder. The surface spectrum of a 3D TI has an odd number of Dirac points, in the vicinity of which electrons are described by the Dirac equation for relativistic massless particles. Strong spin-momentum locking makes the electronic spectrum sensitive to magnetic perturbations. Particularly, exchange coupling between the Dirac electron and a uniform magnetization opens a gap in the surface spectrum and leads to the Anomalous quantum Hall effect~\cite{Haldane1988}. This has a number of interesting manifestations~\cite{Qi2009, Zang2010}, including quantized Faraday and Kerr effects~\cite{Tse2010, Maciejko2010, Efimkin2013}, and has recently been directly observed experimentally~\cite{Chang2013,checkelsky_trajectory_2014}. \\ 
\indent Profound physics emerges when the Dirac electrons couple to spin textures with spatially non-uniform magnetization distributions~\cite{Garate2010, Nomura2010}. A magnetic domain wall in proximity to the Dirac electron generates a chiral state, in analogy to the Jackiw-Rebbi zero mode~\cite{Jackiw1976}, which in turn alters its dynamics and stability~\cite{Tserkovnyak2012, Ferreiros2014, Linder2014, Wickles2012}. One could thus expect more novel properties when the Dirac states of a TI couple to nontrivial spin textures, such as skyrmions~\cite{skyrme_non-linear_1961,ros_sler_spontaneous_2006}, which is the main issue of this work.\\
\begin{figure*}
\centering
\includegraphics[width = \textwidth]{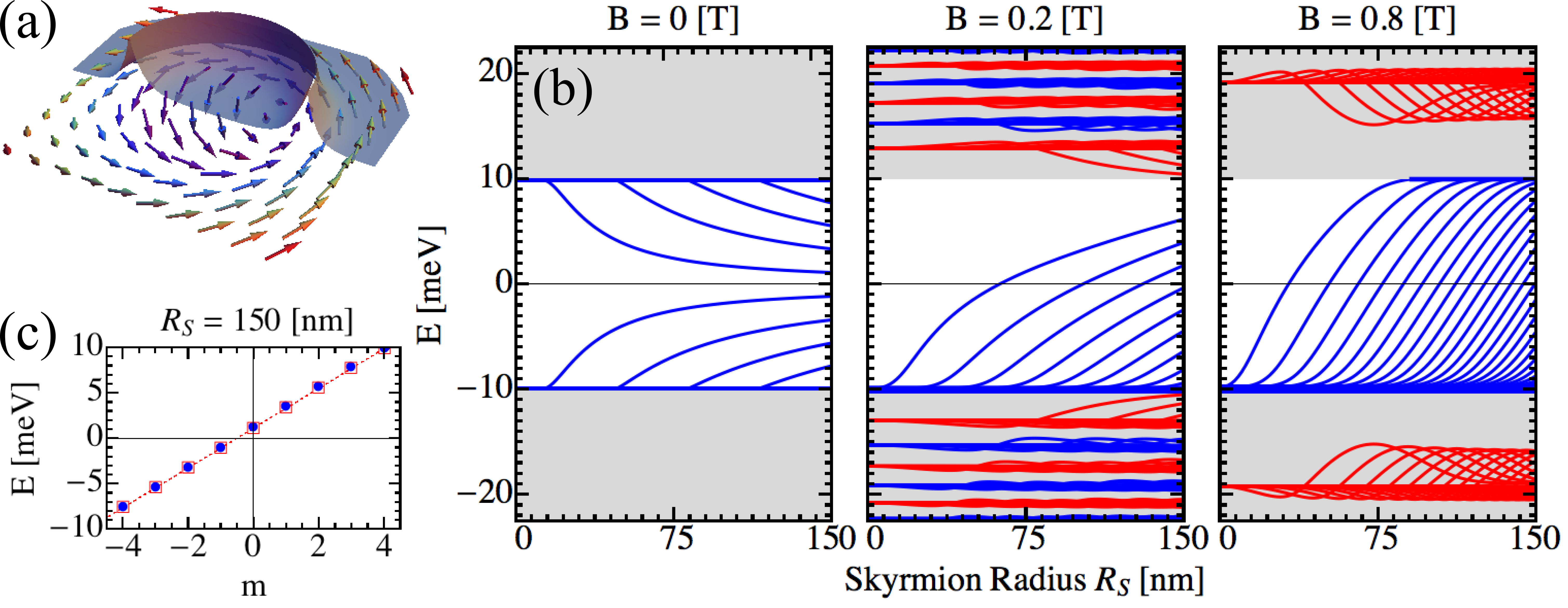}
\caption{(color online)~(a)~Sketch of the wave function $|\Psi(r,\phi)|^2$ (gray) of a TI surface state localized at the skyrmion radius. For clarity we plot $|\Psi|^2$ for $0 < \phi < \pi$ only. The vector field represents the direction of local magnetization $\vec{n}(\vec{r})$ of the skyrmion.~(b)~Electronic spectrum of TI surface proximity coupled to the skyrmion texture as a function of skyrmion radius and magnetic field. Without magnetic field states with orbital quantum number $|m|\leq4$ are presented, with magnetic field $B=0.2~\mathrm{T}$ states with $|m|\leq10$ are presented, and with magnetic field $B=0.8~\mathrm{T}$ states with $|m|\leq25$ are presented. Without magnetic field localized states split from continuous bands $|E|\geq\Delta_\mathrm{S}$ in pairs. In the presence of an external magnetic field localized states split from the zeroth Landau level, which has the energy $E_{0m}=-\Delta_\mathrm{S}-\Delta_\mathrm{Z}$ and is highly degenerate. Higher energy LLs are also shown.~(c)~The energy of bound states as a function of $m$ at large skyrmion radius $R_\mathrm{S}$ without magnetic field.  Numerical solution (blue circles) from equation (\ref{eqn:radial}) matches almost exactly with the semiclassical quantization (red squares) as in equation (\ref{eqn:Bohr}).\label{Fig:RadiusData}}
\end{figure*}
\indent A magnetic skyrmion is a topological spin texture with whirlpool-like structure, illustrated in Fig.~\ref{Fig:RadiusData}-a. Its topology is characterized by the topological invariant 
\begin{equation}
N_\mathrm{S}=\int\frac{d\vec{r}}{4\pi} \vec{n} \left[\frac{\partial \vec{n}}{\partial x}\times \frac{\partial \vec{n}}{\partial y} \right],
\label{eqn:Ns}
\end{equation}
where $\vec{n}(\vec{r})$ is the unit vector describing direction of the magnetization~\cite{rajaraman_solitons_1987}. The skyrmion has been experimentally observed in various chiral magnets~\cite{muhlbauer_skyrmion_2009,yu_real-space_2010} as a result of the the competition between the Dzyaloshinskii-Moriya (DM) interactions, Heisenberg exchange, and Zeeman interaction~\cite{Bogdanov1989, Bogdanov1994, Han2010}. It was first discovered in MnSi in a narrow window at finite temperatures~\cite{muhlbauer_skyrmion_2009}. However, once on a thin film, the skyrmion phase can be greatly extended even down to zero temperature, as long as the fine tuning of the external magnetic field is achieved~\cite{yu_real-space_2010}. Although it has been demonstrated that metallic skyrmions can be driven by the spin transfer torque (STT) from an electric current~\cite{Jonietz2010,Zang2011}, recent appearance of the insulating helimagnet material Cu$_2$OSeO$_3$ raises a more challenging question of how to manipulate insulating skyrmions~\cite{Seki2012,Liu2013,Kong2013,Watanabe2014}. It is important not only from a fundamental point of view, but also for applications in spintronics and memory devices (see Ref.~\cite{Nagaosa2013} for a recent review).\\ 
\indent Here we show that the skyrmion texture in an insulating helimagnet, proximity coupled to TI surface, can become charged due to Dirac surface states localized at the skyrmion radius. The number of localized states can be controlled by an external magnetic field, which is needed for skyrmion phase stabilization, and in the realistic conditions there are a few localized states. Experimentally accessible electric fields can drive individual charged skyrmions in insulating materials at speeds comparable to those seen in metallic systems. Our work shows the first mechanism of manipulating skyrmions in thin films directly via electric field, which opens the door to further investigation of skyrmion manipulation without relying on the STT mechanism.\\
\indent\emph{The electronic spectrum of TI surface states.} In the presence of the skyrmion texture $\vec{n}(\vec{r})$ and external perpendicular magnetic field $B$, corresponding to the vector potential $\vec{A} = B(x \vec{e}_{y}-y \vec{e}_{x})/2$, the surface states of a TI can be described by the Hamiltonian 
\begin{equation}
\mathcal{H} = v_\mathrm{F}\left[\left(\vec{p} -\frac{e}{c}\vec{A}\right) \times \boldsymbol{\sigma}\right]_z -\Delta_\mathrm{S} \vec{n}(\vec{r}) \cdot\boldsymbol{\sigma} -\Delta_\mathrm{Z}\sigma^z.
\label{eqn:H1}
\end{equation}
Here $v_\mathrm{F}$ is the Fermi velocity, $\boldsymbol{\sigma}$ is the vector of Pauli matrices, corresponding to electron's spin, $\Delta_\mathrm{Z} = g\mu_bB$ is the Zeeman shift due to the external field, and $\Delta_\mathrm{S} > 0$ parametrizes ferromagnetic exchange coupling between the thin film and TI surface~\cite{Wei2013}. The out-of plane component of the magnetic texture $\vec{n}_z(\vec{r})$ plays the role of position dependent Dirac mass, while the in-plane component $\vec{n}_{\mathrm{||}}$ can lead to an emergent magnetic field $B_\mathrm{S}(\vec{r}) = c \Delta_\mathrm{S} \mathrm{div} \vec{n}_{\mathrm{||}}(\vec{r})/e \hbar v_\mathrm{F}$. A skyrmion stabilized by DM interaction, as it is for insulating helical magnets, has a solinoidal in-plane magnetization, shown in Fig.~\ref{Fig:RadiusData}-a. Therefore, its magnetic texture can be parametrized by $\vec{n}(\vec{r})=(-\sin \phi \sqrt{1-n_z^2(r)}, \cos \phi \sqrt{1-n_z^2(r)},n_z(r))$. In this case the emergent magnetic field $B_{\mathrm{S}}$ is zero and the in-plane component of magnetic texture can be gauged away. For the out-of-plane component we will use the hard-wall approximation $n_z(r) = 2\Theta(r-R_\mathrm{S}) -1$, where $R_\mathrm{S}$ is the skyrmion radius. The applicability of this approximation is justified below. In materials where skyrmions are stabilized via DM interaction the radius of the skyrmion is not dependent on the external magnetic field. Rather, skyrmion radius depends on the relative strength of the Heisenberg and DM exchange interactions which in turn depends on the material parameters~\cite{Nagaosa2013}. In the following we therefore regard the skyrmion radius as an independent parameter.\\
\indent The Hamiltonian conserves total momentum, and the wave function spinor can be presented as $\Psi(r,\phi)=(F(r)e^{im\phi}/\sqrt{2\pi},G(r)e^{i (m+1)\phi}/\sqrt{2\pi})$, where $m$ is the orbital momentum quantum number connected with the total momentum of Dirac particles as $j = m + \frac{1}{2}$. The radial wave functions $F_m(r)$ and $G_m(r)$ satisfy the equation
\begin{equation*}
\begin{pmatrix}
-\Delta_\mathrm{S}n^\nu_z - \Delta_\mathrm{Z} - E & \frac{v_\mathrm{F}eB}{2c}r - \hbar v_\mathrm{F}\frac{\partial}{\partial r} \\ 
\frac{v_\mathrm{F}eB}{2c}r + \hbar v_\mathrm{F}\frac{\partial}{\partial r} & \Delta_\mathrm{S}n^\nu_z + \Delta_\mathrm{Z} -E\\ 
\end{pmatrix}
\begin{pmatrix}
F^\nu_m\\
G^\nu_m
\end{pmatrix}
\end{equation*}
\begin{equation}
-\begin{pmatrix}
0 & \hbar v_\mathrm{F}\frac{m+1}{r} \\ 
\hbar v_\mathrm{F}\frac{m}{r} & 0 \\ 
\end{pmatrix}
\begin{pmatrix}
F^\nu_m\\
G^\nu_m
\end{pmatrix} = 0 .
\label{eqn:radial}
\end{equation}
Here $\nu=\pm$ for wave function of electrons outside and inside the skyrmion, where $n_z^\pm = \pm1$. Continuity of the Dirac wave function at the skyrmion radius $R_\mathrm{S}$ results in the boundary condition $F^+(R_\mathrm{S})=F^-(R_\mathrm{S})$ and $G^+(R_\mathrm{S})=G^-(R_\mathrm{S})$. The energy at each skyrmion radius is then numerically calculated by solving the transcendental equation $G^-(R_\mathrm{S})/F^-(R_\mathrm{S}) = G^+(R_\mathrm{S})/F^+(R_\mathrm{S})$, generating the spectra shown in Fig.~\ref{Fig:RadiusData}-b. We consider the skyrmion radius and external magnetic field to be separate controlling parameters of the surface state energy. Although skyrmions are usually stabilized in the presence of an external magnetic field, it is instructive to consider at first the electronic spectrum without it.\\
\indent For the following calculations we use $v_\mathrm{F} \approx 0.5\times10^6~\mathrm{m/s}$ and $g_\mathrm{TI} \approx 5$, corresponding to the TI material $\mathrm{Bi}_2\mathrm{Se}_3$, and exchange coupling $\Delta_\mathrm{S} = 10~\mathrm{meV}$. Insulating skyrmions with radius $R_\mathrm{S} \approx 25~\mathrm{nm}$ have been stabilized in $\mathrm{Cu}_2\mathrm{OSeO}_3$ films~\cite{Seki2012, Omrani2014, White2014, White2012} at low temperatures $T<40~\mathrm{K}$ and for magnetic fields  $0.05~\mathrm{T} \leq B \leq 0.2~\mathrm{T}$.\\
\indent \emph{Electronic structure in the absence of the magnetic field.} At $B = 0$ the skyrmion texture does not modify the continuous gapped spectrum $|E| \ge \Delta_\mathrm{S}$~\cite{footnote1}
%
; if the skyrmion radius exceeds the critical radius $R_\mathrm{S}^* = \hbar v_\mathrm{F}/2\Delta_\mathrm{S}$ it can lead to the appearance of localized states as presented in Fig.~\ref{Fig:RadiusData}-b for $B = 0~\mathrm{T}$. The localized states appear in pairs, since the Dirac Hamiltonian possesses electron-hole symmetry $\lbrace H_\mathrm{D}, \hat{\mathrm{K}}\rbrace = 0$ with $\hat{\mathrm{K}} = \sigma^x\mathcal{C}$ where $\mathcal{C}$ is complex conjugation. The states are localized in the vicinity of the skyrmion radius $R_\mathrm{S}$, as presented in Fig.~\ref{Fig:RadiusData}-a with length $\hbar v_\mathrm{F}/\Delta_\mathrm{S}$. These states are well split from the continuous bands and are bound to the skyrmion in case of its motion, with wave functions   
\begin{equation}
\Psi^-_m \propto \begin{bmatrix}
(\epsilon + n^-_z)\mathcal{I}_m(k\rho)\\
\mathcal{I}_{m+1}(k\rho)
\end{bmatrix};\Psi^+_m \propto \begin{bmatrix}
(\epsilon + n^+_z)\mathcal{K}_m(k\rho) \\
\mathcal{K}_{m+1}(k\rho)
\end{bmatrix}
\label{eqn:WFs},
\end{equation}
where $k^2 = n_z^2 -\epsilon^2$, $\epsilon = E/\Delta_\mathrm{S}$, $\rho = \Delta_\mathrm{S}r/\hbar v_\mathrm{F}$, and $\mathcal{I}_m$ ($\mathcal{K}_m$) are modified Bessel functions of the the first (second) kind. \\
\indent The second term in equation (\ref{eqn:radial}) originates from the angular motion and can be neglected if $R_\mathrm{S} \gg 2m R_\mathrm{S}^*,~2(m+1) R_\mathrm{S}^*$. In that case the system of equations (\ref{eqn:radial}) can be reduced to the system of supersymmetric Schr\"{o}dinger equations (SUSY) by squaring $H$ to solve $H^2\Psi_m = E_m^2\Psi_m$, with the SUSY superpotential $W(x) = \Delta_\mathrm{S}n_z(r)$, as it is for a domain wall~\cite{Jackiw1976, Cooper1995}. The SUSY nature of the equations guarantees the presence of a localized state for every angular momentum. Their presence is not sensitive to the shape of the skyrmionic texture $n_z(R)$, so the physics is well captured by the hard-wall anzatz in the $R_\mathrm{S}\gg R_\mathrm{S}^*$ regime. When the the angular motion becomes important, at $R_\mathrm{S}\approx R_\mathrm{S}^*$, the reduction to SUSY is lost and localized states are pushed to the continuum. The energy gain due to the localization becomes smaller than additional kinetic energy due to orbital motion.  \\
\indent For large skyrmions $R_\mathrm{S} \gg R_\mathrm{S}^*$, the energy of bound states can be directly calculated from the semiclassical picture. Along the skyrmion boundary there are chiral states inheriting the dispersion law of massless Dirac particles $\mathrm{E}_\mathrm{\phi}=\hbar v_\mathrm{F} k_\mathrm{\phi}$, and the Bohr-Sommerfeld quantization of their motion is respected; 
\begin{equation}
2\pi R_{\mathrm{S}} k_\mathrm{\phi} +\phi_{\mathrm{B}} = 2 \pi m,
\label{eqn:Bohr}
\end{equation}
where $\phi_\mathrm{B}=-\pi$ is the Berry phase for massless Dirac particles, which appears due to the rotation of the electron's spin when it makes a closed loop in momentum space. This gives $\mathrm{E}_\mathrm{\phi}=\hbar v_\mathrm{F}(m+1/2)/R_\mathrm{S}$ which is in excellent agreement with exact numerical calculation, as presented in Fig.~\ref{Fig:RadiusData}-c. It should be noted that the semiclassical arguments are also weakly dependent on the actual shape of the magnetic texture.\\
\begin{figure}
\centering
\includegraphics[width = 0.5\textwidth]{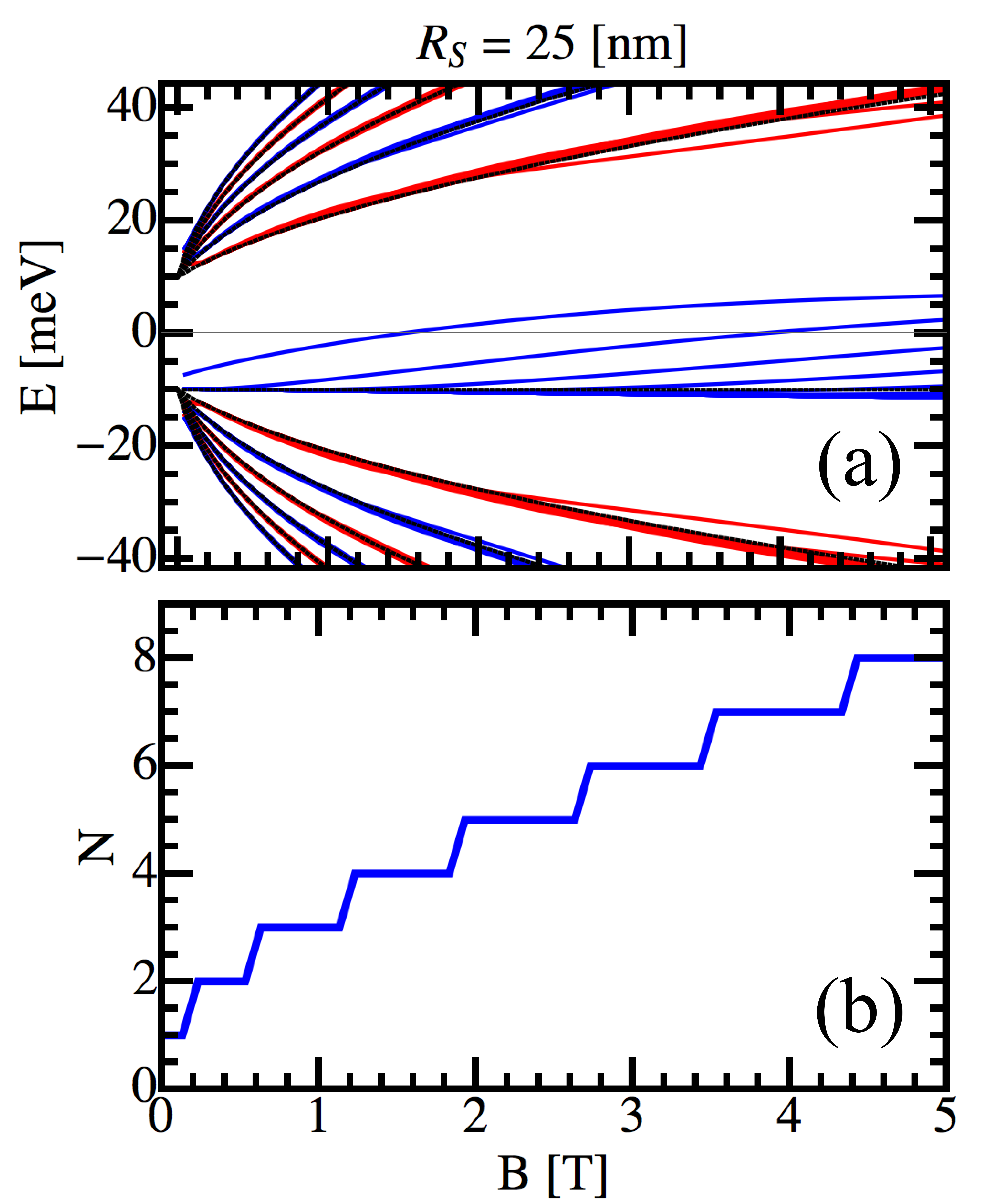}
\caption{(color online)~(a)~Electronic spectrum of TI surface states a function of magnetic field $B$, for skyrmion radius $R_\mathrm{S}=25~\hbox{nm}$. Landau levels with $n \leq 6$ are presented. With increasing magnetic field the number of states split from the zeroth Landau level with the energy $E_{0m}=-\Delta_{\mathrm{S}}-\Delta_{\mathrm{Z}}$ increases. The black lines correspond to the spectrum without the texture.~(b)~Number of split states, bound to the skyrmion, as a function of magnetic field.\label{Fig:F3}}
\end{figure}
\indent \emph{Electronic structure in the presence of the external magnetic field.} Without the skyrmion texture, but in the presence of an external magnetic field, all surface states condensed into Landau Levels (LL) indexed by main and orbital quantum numbers $n$ and $m$, respectively. Landau levels have macroscopic degeneracy $m=\left[0, \Phi/\Phi_0 -1\right]$, where $\Phi$ is the total flux through the TI surface and $\Phi_0=hc/e$ is the flux quantum. In the radial gauge, LL are localized at a cyclotron radius $R_{m} \approx l_\mathrm{B}\sqrt{2(m+1)}$ with magnetic length $l_\mathrm{B} = \sqrt{\hbar c/eB}$. Their energetic spectrum is given by 
\begin{equation}
\begin{split}
E^\pm_{n m} = \pm v_\mathrm{F}\sqrt{\frac{2\hbar^2}{l_\mathrm{B}^2}|n| + \left(\frac{\Delta_\mathrm{Z}}{v_\mathrm{F}}\right)^2} \mbox{~~;~~} n\neq 0,\\
E_{0 m} = -sgn(B)\Delta_\mathrm{Z} \mbox{~~;~~} n = 0 . 
\label{eqn:E0}
\end{split}
\end{equation}
The zeroth LL is spin polarized due to the orbital effect of magnetic field and its shift is sensitive to Zeeman coupling $\Delta_\mathrm{Z}$, while other LLs are not sensitive to its sign.\\
\indent In the presence of the magnetic field and the skyrmion texture, discrete states appear which are well split from the macroscopically degenerate $n=0$ LL. For these states, the cyclotron radius $R_m$ approximately coincides with the skyrmion radius $R_\mathrm{S}$, therefore the states are localized at the skyrmion boundary and move with the skyrmion. The zeroth LL states with $R_m\gg R_\mathrm{S}$ approach $E_{0 m} = -\mathrm{sgn}(B)(\Delta_\mathrm{Z} + \Delta_\mathrm{S})$. These states are weakly affected by the skyrmion texture and are not correlated with its motion, in analogy to the delocalized states in the absence of a magnetic field. The energetic spectrum of TI surface states as function of skyrmion radius is presented in Fig.~\ref{Fig:RadiusData}-b for magnetic field values  $B = 0.2~\mathrm{T}$ and $0.8~\mathrm{T}$. States bound to the skyrmion no longer appear in pairs since electron-hole symmetry is broken by the magnetic field. States with $|E|\geq \Delta_\mathrm{S}$ condense into LLs with $n \neq 0$ which are weakly affected by the skyrmion texture. States localized at the skyrmion boundary have the wave functions 
\begin{equation}
\begin{split}
\Psi^-_{m} \propto \rho^{m}e^{-\frac{\rho^2b}{2}}\begin{bmatrix}
F_1(\alpha^-, 1+m; \rho^2b)\\
\rho F_1(\alpha^- +1, 2+m; \rho^2b)
\end{bmatrix} ; \\
\Psi^+_{m} \propto \rho^{m}e^{-\frac{\rho^2b}{2}}\begin{bmatrix}
\mathcal{U}(\alpha^+, 1+m; \rho^2b)\\
\rho \mathcal{U}(\alpha^++1, 2+m; \rho^2b)
\end{bmatrix},
\end{split}
\end{equation}
where $F_1$, $\mathcal{U}$ are confluent hypergeometric functions of the first and second kind, parameterized by $\alpha^{\pm} = ((n_z^\pm +\delta)^2 - \epsilon^2)/4b$. $b = B/B_0$ where $B_0 = 2\Delta_\mathrm{S}^2c/\hbar v_\mathrm{F}^2e$, $\delta = \Delta_\mathrm{Z}/\Delta_\mathrm{S}$ and $\rho = \Delta_\mathrm{S} r/\hbar v_\mathrm{F}$.\\
\indent The dependence of the TI surface spectrum as a function of magnetic field $B$ for $R_\mathrm{S}=25~\hbox{nm}$ is presented in Fig.~\ref{Fig:F3}-a. The presence of states bound to the skyrmion is robust to the external magnetic field. Moreover due to the interplay between LL and skyrmion confinement the number of bound states increases with magnetic field as presented in Fig.~\ref{Fig:F3}-b. The appearance of bound states can lead to charging of the skyrmion texture. The electric charge is $eN_\mathrm{B}$, where $N_\mathrm{B}$ is the number of additional occupied electronic states in comparison to the TI surface without the skyrmion texture. The value $N_\mathrm{B}$ depends on local chemical potential and can be controlled by magnetic field. Our mechanism of the texture charging differs from the one in Ref.~\cite{Nomura2010} based on quantized response of electronic density to the emergent magnetic field $B_\mathrm{S}(\vec{r})$, which originates from a magnetic texture but is zero $B_\mathrm{S}(\vec{r}) = 0$ in our case.\\
\indent The radius of skyrmions in Cu$_2$OSeO$_3$, $R_\mathrm{S}=25~\hbox{nm}$, exceeds the critical radius $R^*_\mathrm{S}= \hbar v_\mathrm{F}/2\Delta_\mathrm{S} \approx 16~\mathrm{nm}$ and there is one bound state for magnetic field $0.05~\hbox{T}\leq B \leq 0.2 ~\hbox{T}$. The skyrmionic charge can be equal to the single charge of an electron. If the radius of individual skyrmions is larger, the skyrmion phase is stabilized in a wider magnetic field interval, or for greater values of exchange coupling $\Delta_\mathrm{S}$, the skyrmion will host a few bound states and the skyrmion charge becomes more tunable by magnetic field or doping. \\
\indent\emph{Dynamics of charged skyrmions.} Being charged particle-like objects, skyrmions with bound surface states can be described by the following Lagrangian
\begin{equation}
\mathcal{L}= Q_\mathrm{S}\left[\dot{\vec{R}}\times \vec{R}\right]_z  + e N_\mathrm{B}\left( \phi(\vec{R})-\frac{\dot{R}}{c} A(\vec{R})\right)+ \frac{m_\mathrm{S} \dot{R}^2}{2} . 
\label{eqn:Lag}
\end{equation}
Here the first term represents the Magnus force originating from spin dynamics~\cite{Klauder1979, Stone1989, Stone1996}, and $Q_\mathrm{S}=4\pi \rho_\mathrm{S} S \hbar N_\mathrm{S}$ where $N_\mathrm{S}$ is the skyrmion topological invariant, with $\rho_\mathrm{S}$ and $S$ the density and amplitude of spins in the skyrmionic magnet. The second term represents the interaction of skyrmion with external electromagnetic field, while the last term corresponds to the possible additional kinetic energy with phenomenologically induced skyrmion mass $m_\mathrm{S}$. In the presence of an electric field $\vec{E} = (E_x, 0)$ the skyrmion acquires the velocity
\begin{align}
\dot{R}_y &= -\frac{eN_\mathrm{B}c\left(2Q_\mathrm{S}c+eN_\mathrm{B}B\right)}{\left(2Q_\mathrm{S}c+eN_\mathrm{B}B\right)^2 + c^2\Gamma^2}E_x ; \\ \dot{R}_x &= -\frac{eN_\mathrm{B}c^2\Gamma}{\left(2Q_\mathrm{S}c+eN_\mathrm{B}B\right)^2 + c^2\Gamma^2}E_x . 
\end{align} 
Here $\Gamma$ is a phenomenological friction parameter. Note that the mass term in equation (\ref{eqn:Lag}) only provides the initial acceleration of the skyrmion. However the skyrmion's final steady velocity is determined from the balance between electric field and damping, thus it is independent of the mass $m_\mathrm{S}$. In the limit $\Gamma \rightarrow 0$, the system exhibits Hall motion where $\dot{R}_{x} = 0$ and $\dot{R}_{y} = -\mu_\mathrm{S}  E_x$, defining skyrmion mobility $\mu_\mathrm{S} = eN_\mathrm{B}c/ \left(2Q_\mathrm{S}c+ eN_\mathrm{B}B\right)$. Taking $N_\mathrm{B} = 1$, $B = 0.2~\mathrm{T}$, $\rho \sim d/a^3$, where $d \approx 100~\mathrm{nm}$ is the film thickness and $a \sim 8.9~\mathrm{\AA}$ is the crystallographic lattice constant, we have $\mu_\mathrm{S} \sim 1\times10^{-6}~\mathrm{m^2/V s}$. Electric fields as low as $10^2~\mathrm{V/m}$ can induce drift velocities $v_\mathrm{H} \sim 0.1~\mathrm{mm/s}$, comparable to skyrmions driven by conduction electrons in metallic systems~\cite{Schulz2012}, according to the STT mechanism~ (Ref.~\cite{Nagaosa2013,Jonietz2010,Everschor2011,Everschor2012,Zang2011} and references therein). In the system under consideration the skyrmionic material is insulating and the the surface spectrum of TI is gapped away from the skyrmion, which makes the STT mechanism unimportant. \\ 
\indent It has recently been proposed that insulating skyrmions can be driven by a temperature gradient~\cite{Kong2013}. For electric fields $E\sim 10^5$~V/m or greater the velocity of a skyrmion driven by electric field is higher than the expected velocity due to thermal gradient ($0.1~\mathrm{m/s}$ in ref.~\cite{Kong2013}) and, more importantly, an electric field is easier to manipulate than a temperature gradient. It has also been shown that in multiferroic materials, which include Cu$_2$OSeO$_3$, skyrmion textures have an intrinsic dipole moment which enables them to couple to to the gradient of electric field~\cite{White2012, Seki2012_2,Liu2013, White2014}. Regardless of the difficulty of applying a gradient in electric field, the estimated Hall velocity $v_\mathrm{H}$ in our approach is two orders of magnitude greater than $v_\mathrm{H}$ due to magnetoelectric coupling~\cite{Liu2013}, where $v_\mathrm{H}\approx(\lambda R_\mathrm{S} /2 \pi S \hbar) \Delta\mathrm{E}\sim10^{-3}~\mathrm{mm/s}$ with $R_\mathrm{S} = 25~\mathrm{nm}$, dipolar coupling $\lambda\sim10^{-33}\approx\mathrm{C\cdot m}$~\cite{Omrani2014}, and field strength \emph{difference} $\Delta\mathrm{E} \sim 10^2~\mathrm{V/m}$. We therefore conclude that our mechanism is very effective due to the direct coupling of skyrmions with electric field.\\  
\indent In an experiment one can use a TI such as Bi$_2$Se$_3$ as a substrate and epitaxially grow the Cu$_2$OSeO$_3$ thin film on top of it. An ultra-thin buffer layer inserted between these two can help to avoid lattice mismatch while keeping the coupling between Dirac electrons and the magnetization. Concerning the experimental difficulty in growing Cu$_2$OSeO$_3$ films, one can also use monolayers of ferromagnet instead, which is generally insulating. The interfacial DM interaction can also generate skyrmions therein~\cite{heinze_spontaneous_2011}. The whole sample is sandwiched in a pair of electrodes, which generates an electric field.\\
\indent It should be noted that the skyrmion phase can be intrinsic to surface states of TI in the presence of strong Coulomb repulsion and hexagonal warping, which takes place at high doping~\cite{Baum2012}. In that regime our mechanism is not of importance, since the skyrmion texture changes the surface spectrum in the vicinity of the Dirac point which is deeply buried under the Fermi level. \\ 
\indent We have focused on the behavior of a single charged skyrmion. Skyrmions most frequently appear in a lattice which can be pinned by disorder and the underlying atomic lattice. Therefore in an experiment this effect would manifest as a rotation of the skyrmion lattice caused by Hall motion of individual charged skyrmions. We predict that the angle of skyrmion lattice rotation will be proportional to the electric field and the number of bound states at the skyrmion radius. The angle of rotation should change within the skyrmion phase in response to increased DC electric field. To rule out magnetoelectric effects, an experiment at constant electric field but varying magnetic field could be conducted, provided that the stabilizing magnetic field range of skyrmions is big enough to tune the number of bound states. In this case we predict a change in rotation angle in response to an increase in magnetic field, indicating additional bound surface states.\\
\indent This research was supported by the U.S. Department of Energy DOE-BES (Grant No. DE- 317 SC0001911) and the Simons Foundation (D.E. and V.G). H.H. acknowledges support from the National Science Foundation CAREER grant DMR-0847224 and additional fellowship support from the National Physical Science Consortium and NSA. J.Z. was supported by the Theoretical Interdisciplinary Physics and Astrophysics Center, the U.S. Department of Energy under Award DEFG02-08ER46544, and the National Science Foundation under Grant No. ECCS-1408168.

\bibliographystyle{apsrev}
\bibliography{main}

\begin{thebibliography}{47}
\expandafter\ifx\csname natexlab\endcsname\relax\def\natexlab#1{#1}\fi
\expandafter\ifx\csname bibnamefont\endcsname\relax
  \def\bibnamefont#1{#1}\fi
\expandafter\ifx\csname bibfnamefont\endcsname\relax
  \def\bibfnamefont#1{#1}\fi
\expandafter\ifx\csname citenamefont\endcsname\relax
  \def\citenamefont#1{#1}\fi
\expandafter\ifx\csname url\endcsname\relax
  \def\url#1{\texttt{#1}}\fi
\expandafter\ifx\csname urlprefix\endcsname\relax\def\urlprefix{URL }\fi
\providecommand{\bibinfo}[2]{#2}
\providecommand{\eprint}[2][]{\url{#2}}

\bibitem[{\citenamefont{Hasan and Kane}(2010)}]{Hasan2010}
\bibinfo{author}{\bibfnamefont{M.~Z.} \bibnamefont{Hasan}} \bibnamefont{and}
  \bibinfo{author}{\bibfnamefont{C.~L.} \bibnamefont{Kane}},
  \bibinfo{journal}{Rev. Mod. Phys.} \textbf{\bibinfo{volume}{82}},
  \bibinfo{pages}{3045} (\bibinfo{year}{2010}).

\bibitem[{\citenamefont{Qi and Zhang}(2011)}]{Qi2011}
\bibinfo{author}{\bibfnamefont{X.-L.} \bibnamefont{Qi}} \bibnamefont{and}
  \bibinfo{author}{\bibfnamefont{S.-C.} \bibnamefont{Zhang}},
  \bibinfo{journal}{Rev. Mod. Phys.} \textbf{\bibinfo{volume}{83}},
  \bibinfo{pages}{1057} (\bibinfo{year}{2011}).

\bibitem[{\citenamefont{Haldane}(1988)}]{Haldane1988}
\bibinfo{author}{\bibfnamefont{F.~D.~M.} \bibnamefont{Haldane}},
  \bibinfo{journal}{Phys. Rev. Lett.} \textbf{\bibinfo{volume}{61}},
  \bibinfo{pages}{2015} (\bibinfo{year}{1988}).

\bibitem[{\citenamefont{Qi et~al.}(2009)\citenamefont{Qi, Li, Zang, and
  Zhang}}]{Qi2009}
\bibinfo{author}{\bibfnamefont{X.-L.} \bibnamefont{Qi}},
  \bibinfo{author}{\bibfnamefont{R.}~\bibnamefont{Li}},
  \bibinfo{author}{\bibfnamefont{J.}~\bibnamefont{Zang}}, \bibnamefont{and}
  \bibinfo{author}{\bibfnamefont{S.-C.} \bibnamefont{Zhang}},
  \bibinfo{journal}{Science} \textbf{\bibinfo{volume}{323}},
  \bibinfo{pages}{1184} (\bibinfo{year}{2009}).

\bibitem[{\citenamefont{Zang and Nagaosa}(2010)}]{Zang2010}
\bibinfo{author}{\bibfnamefont{J.}~\bibnamefont{Zang}} \bibnamefont{and}
  \bibinfo{author}{\bibfnamefont{N.}~\bibnamefont{Nagaosa}},
  \bibinfo{journal}{Phys. Rev. B} \textbf{\bibinfo{volume}{81}},
  \bibinfo{pages}{245125} (\bibinfo{year}{2010}).

\bibitem[{\citenamefont{Tse and MacDonald}(2010)}]{Tse2010}
\bibinfo{author}{\bibfnamefont{W.-K.} \bibnamefont{Tse}} \bibnamefont{and}
  \bibinfo{author}{\bibfnamefont{A.~H.} \bibnamefont{MacDonald}},
  \bibinfo{journal}{Phys. Rev. Lett.} \textbf{\bibinfo{volume}{105}},
  \bibinfo{pages}{057401} (\bibinfo{year}{2010}).

\bibitem[{\citenamefont{Maciejko et~al.}(2010)\citenamefont{Maciejko, Qi, Drew,
  and Zhang}}]{Maciejko2010}
\bibinfo{author}{\bibfnamefont{J.}~\bibnamefont{Maciejko}},
  \bibinfo{author}{\bibfnamefont{X.-L.} \bibnamefont{Qi}},
  \bibinfo{author}{\bibfnamefont{H.~D.} \bibnamefont{Drew}}, \bibnamefont{and}
  \bibinfo{author}{\bibfnamefont{S.-C.} \bibnamefont{Zhang}},
  \bibinfo{journal}{Phys. Rev. Lett.} \textbf{\bibinfo{volume}{105}},
  \bibinfo{pages}{166803} (\bibinfo{year}{2010}).

\bibitem[{\citenamefont{Efimkin and Lozovik}(2013)}]{Efimkin2013}
\bibinfo{author}{\bibfnamefont{D.}~\bibnamefont{Efimkin}} \bibnamefont{and}
  \bibinfo{author}{\bibfnamefont{Y.~E.} \bibnamefont{Lozovik}},
  \bibinfo{journal}{Phys. Rev. B} \textbf{\bibinfo{volume}{87}},
  \bibinfo{pages}{245416} (\bibinfo{year}{2013}).

\bibitem[{\citenamefont{Chang et~al.}(2013)\citenamefont{Chang, Zhang, Feng,
  Shen, Zhang, Guo, Li, Ou, Wei, Wang et~al.}}]{Chang2013}
\bibinfo{author}{\bibfnamefont{C.-Z.} \bibnamefont{Chang}},
  \bibinfo{author}{\bibfnamefont{J.}~\bibnamefont{Zhang}},
  \bibinfo{author}{\bibfnamefont{X.}~\bibnamefont{Feng}},
  \bibinfo{author}{\bibfnamefont{J.}~\bibnamefont{Shen}},
  \bibinfo{author}{\bibfnamefont{Z.}~\bibnamefont{Zhang}},
  \bibinfo{author}{\bibfnamefont{M.}~\bibnamefont{Guo}},
  \bibinfo{author}{\bibfnamefont{K.}~\bibnamefont{Li}},
  \bibinfo{author}{\bibfnamefont{Y.}~\bibnamefont{Ou}},
  \bibinfo{author}{\bibfnamefont{P.}~\bibnamefont{Wei}},
  \bibinfo{author}{\bibfnamefont{L.-L.} \bibnamefont{Wang}},
  \bibnamefont{et~al.}, \bibinfo{journal}{Science}
  \textbf{\bibinfo{volume}{340}}, \bibinfo{pages}{167} (\bibinfo{year}{2013}).

\bibitem[{\citenamefont{Checkelsky et~al.}(2014)\citenamefont{Checkelsky,
  Yoshimi, Tsukazaki, Takahashi, Kozuka, Falson, Kawasaki, and
  Tokura}}]{checkelsky_trajectory_2014}
\bibinfo{author}{\bibfnamefont{J.~G.} \bibnamefont{Checkelsky}},
  \bibinfo{author}{\bibfnamefont{R.}~\bibnamefont{Yoshimi}},
  \bibinfo{author}{\bibfnamefont{A.}~\bibnamefont{Tsukazaki}},
  \bibinfo{author}{\bibfnamefont{K.~S.} \bibnamefont{Takahashi}},
  \bibinfo{author}{\bibfnamefont{Y.}~\bibnamefont{Kozuka}},
  \bibinfo{author}{\bibfnamefont{J.}~\bibnamefont{Falson}},
  \bibinfo{author}{\bibfnamefont{M.}~\bibnamefont{Kawasaki}}, \bibnamefont{and}
  \bibinfo{author}{\bibfnamefont{Y.}~\bibnamefont{Tokura}},
  \bibinfo{journal}{Nat. Phys.} \textbf{\bibinfo{volume}{10}},
  \bibinfo{pages}{731} (\bibinfo{year}{2014}).

\bibitem[{\citenamefont{Garate and Franz}(2010)}]{Garate2010}
\bibinfo{author}{\bibfnamefont{I.}~\bibnamefont{Garate}} \bibnamefont{and}
  \bibinfo{author}{\bibfnamefont{M.}~\bibnamefont{Franz}},
  \bibinfo{journal}{Phys. Rev. Lett.} \textbf{\bibinfo{volume}{104}},
  \bibinfo{pages}{146802} (\bibinfo{year}{2010}).

\bibitem[{\citenamefont{Nomura and Nagaosa}(2010)}]{Nomura2010}
\bibinfo{author}{\bibfnamefont{K.}~\bibnamefont{Nomura}} \bibnamefont{and}
  \bibinfo{author}{\bibfnamefont{N.}~\bibnamefont{Nagaosa}},
  \bibinfo{journal}{Phys. Rev. B} \textbf{\bibinfo{volume}{82}},
  \bibinfo{pages}{161401} (\bibinfo{year}{2010}).

\bibitem[{\citenamefont{Jackiw and Rebbi}(1976)}]{Jackiw1976}
\bibinfo{author}{\bibfnamefont{R.}~\bibnamefont{Jackiw}} \bibnamefont{and}
  \bibinfo{author}{\bibfnamefont{C.}~\bibnamefont{Rebbi}},
  \bibinfo{journal}{Phys. Rev. D} \textbf{\bibinfo{volume}{13}},
  \bibinfo{pages}{3398} (\bibinfo{year}{1976}).

\bibitem[{\citenamefont{Tserkovnyak and Loss}(2012)}]{Tserkovnyak2012}
\bibinfo{author}{\bibfnamefont{Y.}~\bibnamefont{Tserkovnyak}} \bibnamefont{and}
  \bibinfo{author}{\bibfnamefont{D.}~\bibnamefont{Loss}},
  \bibinfo{journal}{Phys. Rev. Lett.} \textbf{\bibinfo{volume}{108}},
  \bibinfo{pages}{187201} (\bibinfo{year}{2012}).

\bibitem[{\citenamefont{Ferreiros and Cortijo}(2014)}]{Ferreiros2014}
\bibinfo{author}{\bibfnamefont{Y.}~\bibnamefont{Ferreiros}} \bibnamefont{and}
  \bibinfo{author}{\bibfnamefont{A.}~\bibnamefont{Cortijo}},
  \bibinfo{journal}{Phys. Rev. B} \textbf{\bibinfo{volume}{89}},
  \bibinfo{pages}{024413} (\bibinfo{year}{2014}).

\bibitem[{\citenamefont{Linder}(2014)}]{Linder2014}
\bibinfo{author}{\bibfnamefont{J.}~\bibnamefont{Linder}},
  \bibinfo{journal}{Phys. Rev. B} \textbf{\bibinfo{volume}{90}},
  \bibinfo{pages}{041412} (\bibinfo{year}{2014}).

\bibitem[{\citenamefont{Wickles and Belzig}(2012)}]{Wickles2012}
\bibinfo{author}{\bibfnamefont{C.}~\bibnamefont{Wickles}} \bibnamefont{and}
  \bibinfo{author}{\bibfnamefont{W.}~\bibnamefont{Belzig}},
  \bibinfo{journal}{Phys. Rev. B} \textbf{\bibinfo{volume}{86}},
  \bibinfo{pages}{035151} (\bibinfo{year}{2012}).

\bibitem[{\citenamefont{Skyrme}(1961)}]{skyrme_non-linear_1961}
\bibinfo{author}{\bibfnamefont{T.~H.~R.} \bibnamefont{Skyrme}},
  \bibinfo{journal}{Proc. of the Royal Soc. of London.}
  \textbf{\bibinfo{volume}{260}}, \bibinfo{pages}{127} (\bibinfo{year}{1961}).

\bibitem[{\citenamefont{R{\"o}{\ss}ler
  et~al.}(2006)\citenamefont{R{\"o}{\ss}ler, Bogdanov, and
  Pfleiderer}}]{ros_sler_spontaneous_2006}
\bibinfo{author}{\bibfnamefont{U.}~\bibnamefont{R{\"o}{\ss}ler}},
  \bibinfo{author}{\bibfnamefont{A.}~\bibnamefont{Bogdanov}}, \bibnamefont{and}
  \bibinfo{author}{\bibfnamefont{C.}~\bibnamefont{Pfleiderer}},
  \bibinfo{journal}{Nature} \textbf{\bibinfo{volume}{442}},
  \bibinfo{pages}{797} (\bibinfo{year}{2006}).

\bibitem[{\citenamefont{Rajaraman}(1987)}]{rajaraman_solitons_1987}
\bibinfo{author}{\bibfnamefont{R.}~\bibnamefont{Rajaraman}},
  \emph{\bibinfo{title}{Solitons and Instantons}}
  (\bibinfo{publisher}{North-Holland}, \bibinfo{address}{Amsterdam},
  \bibinfo{year}{1987}).

\bibitem[{\citenamefont{M{\"u}hlbauer et~al.}(2009)\citenamefont{M{\"u}hlbauer,
  Binz, Jonietz, Pfleiderer, Rosch, Neubauer, Georgii, and
  B{\"o}ni}}]{muhlbauer_skyrmion_2009}
\bibinfo{author}{\bibfnamefont{S.}~\bibnamefont{M{\"u}hlbauer}},
  \bibinfo{author}{\bibfnamefont{B.}~\bibnamefont{Binz}},
  \bibinfo{author}{\bibfnamefont{F.}~\bibnamefont{Jonietz}},
  \bibinfo{author}{\bibfnamefont{C.}~\bibnamefont{Pfleiderer}},
  \bibinfo{author}{\bibfnamefont{A.}~\bibnamefont{Rosch}},
  \bibinfo{author}{\bibfnamefont{A.}~\bibnamefont{Neubauer}},
  \bibinfo{author}{\bibfnamefont{R.}~\bibnamefont{Georgii}}, \bibnamefont{and}
  \bibinfo{author}{\bibfnamefont{P.}~\bibnamefont{B{\"o}ni}},
  \bibinfo{journal}{Science} \textbf{\bibinfo{volume}{323}},
  \bibinfo{pages}{915} (\bibinfo{year}{2009}).

\bibitem[{\citenamefont{Yu et~al.}(2010)\citenamefont{Yu, Onose, Kanazawa,
  Park, Han, Matsui, Nagaosa, and Tokura}}]{yu_real-space_2010}
\bibinfo{author}{\bibfnamefont{X.~Z.} \bibnamefont{Yu}},
  \bibinfo{author}{\bibfnamefont{Y.}~\bibnamefont{Onose}},
  \bibinfo{author}{\bibfnamefont{N.}~\bibnamefont{Kanazawa}},
  \bibinfo{author}{\bibfnamefont{J.~H.} \bibnamefont{Park}},
  \bibinfo{author}{\bibfnamefont{J.~H.} \bibnamefont{Han}},
  \bibinfo{author}{\bibfnamefont{Y.}~\bibnamefont{Matsui}},
  \bibinfo{author}{\bibfnamefont{N.}~\bibnamefont{Nagaosa}}, \bibnamefont{and}
  \bibinfo{author}{\bibfnamefont{Y.}~\bibnamefont{Tokura}},
  \bibinfo{journal}{Nature} \textbf{\bibinfo{volume}{465}},
  \bibinfo{pages}{901} (\bibinfo{year}{2010}).

\bibitem[{\citenamefont{Bogdanov and Yablonskii}(1989)}]{Bogdanov1989}
\bibinfo{author}{\bibfnamefont{A.}~\bibnamefont{Bogdanov}} \bibnamefont{and}
  \bibinfo{author}{\bibfnamefont{D.}~\bibnamefont{Yablonskii}},
  \bibinfo{journal}{Zh. Eksp. Teor. Fiz} \textbf{\bibinfo{volume}{95}},
  \bibinfo{pages}{182} (\bibinfo{year}{1989}).

\bibitem[{\citenamefont{Bogdanov and Hubert}(1994)}]{Bogdanov1994}
\bibinfo{author}{\bibfnamefont{A.}~\bibnamefont{Bogdanov}} \bibnamefont{and}
  \bibinfo{author}{\bibfnamefont{A.}~\bibnamefont{Hubert}},
  \bibinfo{journal}{J. of Magnetism and Magnetic Mat.}
  \textbf{\bibinfo{volume}{138}}, \bibinfo{pages}{255} (\bibinfo{year}{1994}).

\bibitem[{\citenamefont{Han et~al.}(2010)\citenamefont{Han, Zang, Yang, Park,
  and Nagaosa}}]{Han2010}
\bibinfo{author}{\bibfnamefont{J.~H.} \bibnamefont{Han}},
  \bibinfo{author}{\bibfnamefont{J.}~\bibnamefont{Zang}},
  \bibinfo{author}{\bibfnamefont{Z.}~\bibnamefont{Yang}},
  \bibinfo{author}{\bibfnamefont{J.-H.} \bibnamefont{Park}}, \bibnamefont{and}
  \bibinfo{author}{\bibfnamefont{N.}~\bibnamefont{Nagaosa}},
  \bibinfo{journal}{Phys. Rev. B} \textbf{\bibinfo{volume}{82}},
  \bibinfo{pages}{094429} (\bibinfo{year}{2010}).

\bibitem[{\citenamefont{Jonietz et~al.}(2010)\citenamefont{Jonietz,
  M{\"u}hlbauer, Pfleiderer, Neubauer, M{\"u}nzer, Bauer, Adams, Georgii,
  B{\"o}ni, Duine et~al.}}]{Jonietz2010}
\bibinfo{author}{\bibfnamefont{F.}~\bibnamefont{Jonietz}},
  \bibinfo{author}{\bibfnamefont{S.}~\bibnamefont{M{\"u}hlbauer}},
  \bibinfo{author}{\bibfnamefont{C.}~\bibnamefont{Pfleiderer}},
  \bibinfo{author}{\bibfnamefont{A.}~\bibnamefont{Neubauer}},
  \bibinfo{author}{\bibfnamefont{W.}~\bibnamefont{M{\"u}nzer}},
  \bibinfo{author}{\bibfnamefont{A.}~\bibnamefont{Bauer}},
  \bibinfo{author}{\bibfnamefont{T.}~\bibnamefont{Adams}},
  \bibinfo{author}{\bibfnamefont{R.}~\bibnamefont{Georgii}},
  \bibinfo{author}{\bibfnamefont{P.}~\bibnamefont{B{\"o}ni}},
  \bibinfo{author}{\bibfnamefont{R.}~\bibnamefont{Duine}},
  \bibnamefont{et~al.}, \bibinfo{journal}{Science}
  \textbf{\bibinfo{volume}{330}}, \bibinfo{pages}{1648} (\bibinfo{year}{2010}).

\bibitem[{\citenamefont{Zang et~al.}(2011)\citenamefont{Zang, Mostovoy, Han,
  and Nagaosa}}]{Zang2011}
\bibinfo{author}{\bibfnamefont{J.}~\bibnamefont{Zang}},
  \bibinfo{author}{\bibfnamefont{M.}~\bibnamefont{Mostovoy}},
  \bibinfo{author}{\bibfnamefont{J.~H.} \bibnamefont{Han}}, \bibnamefont{and}
  \bibinfo{author}{\bibfnamefont{N.}~\bibnamefont{Nagaosa}},
  \bibinfo{journal}{Phys. Rev. Lett.} \textbf{\bibinfo{volume}{107}},
  \bibinfo{pages}{136804} (\bibinfo{year}{2011}).

\bibitem[{\citenamefont{Seki et~al.}(2012{\natexlab{a}})\citenamefont{Seki, Yu,
  Ishiwata, and Tokura}}]{Seki2012}
\bibinfo{author}{\bibfnamefont{S.}~\bibnamefont{Seki}},
  \bibinfo{author}{\bibfnamefont{X.}~\bibnamefont{Yu}},
  \bibinfo{author}{\bibfnamefont{S.}~\bibnamefont{Ishiwata}}, \bibnamefont{and}
  \bibinfo{author}{\bibfnamefont{Y.}~\bibnamefont{Tokura}},
  \bibinfo{journal}{Science} \textbf{\bibinfo{volume}{336}},
  \bibinfo{pages}{198} (\bibinfo{year}{2012}{\natexlab{a}}).

\bibitem[{\citenamefont{Liu et~al.}(2013)\citenamefont{Liu, Li, and
  Han}}]{Liu2013}
\bibinfo{author}{\bibfnamefont{Y.-H.} \bibnamefont{Liu}},
  \bibinfo{author}{\bibfnamefont{Y.-Q.} \bibnamefont{Li}}, \bibnamefont{and}
  \bibinfo{author}{\bibfnamefont{J.~H.} \bibnamefont{Han}},
  \bibinfo{journal}{Phys. Rev. B} \textbf{\bibinfo{volume}{87}},
  \bibinfo{pages}{100402} (\bibinfo{year}{2013}).

\bibitem[{\citenamefont{Kong and Zang}(2013)}]{Kong2013}
\bibinfo{author}{\bibfnamefont{L.}~\bibnamefont{Kong}} \bibnamefont{and}
  \bibinfo{author}{\bibfnamefont{J.}~\bibnamefont{Zang}},
  \bibinfo{journal}{Phys. Rev. Lett.} \textbf{\bibinfo{volume}{111}},
  \bibinfo{pages}{067203} (\bibinfo{year}{2013}).

\bibitem[{\citenamefont{Watanabe and Vishwanath}(2014)}]{Watanabe2014}
\bibinfo{author}{\bibfnamefont{H.}~\bibnamefont{Watanabe}} \bibnamefont{and}
  \bibinfo{author}{\bibfnamefont{A.}~\bibnamefont{Vishwanath}},
  \bibinfo{journal}{arXiv preprint arXiv:1410.2213}  (\bibinfo{year}{2014}).

\bibitem[{\citenamefont{Nagaosa and Tokura}(2013)}]{Nagaosa2013}
\bibinfo{author}{\bibfnamefont{N.}~\bibnamefont{Nagaosa}} \bibnamefont{and}
  \bibinfo{author}{\bibfnamefont{Y.}~\bibnamefont{Tokura}},
  \bibinfo{journal}{Nat. Nano.} \textbf{\bibinfo{volume}{8}},
  \bibinfo{pages}{899} (\bibinfo{year}{2013}).

\bibitem[{\citenamefont{Wei et~al.}(2013)\citenamefont{Wei, Katmis, Assaf,
  Steinberg, Jarillo-Herrero, Heiman, and Moodera}}]{Wei2013}
\bibinfo{author}{\bibfnamefont{P.}~\bibnamefont{Wei}},
  \bibinfo{author}{\bibfnamefont{F.}~\bibnamefont{Katmis}},
  \bibinfo{author}{\bibfnamefont{B.~A.} \bibnamefont{Assaf}},
  \bibinfo{author}{\bibfnamefont{H.}~\bibnamefont{Steinberg}},
  \bibinfo{author}{\bibfnamefont{P.}~\bibnamefont{Jarillo-Herrero}},
  \bibinfo{author}{\bibfnamefont{D.}~\bibnamefont{Heiman}}, \bibnamefont{and}
  \bibinfo{author}{\bibfnamefont{J.~S.} \bibnamefont{Moodera}},
  \bibinfo{journal}{Phys. Rev. Lett.} \textbf{\bibinfo{volume}{110}},
  \bibinfo{pages}{186807} (\bibinfo{year}{2013}).

\bibitem[{\citenamefont{Omrani et~al.}(2014)\citenamefont{Omrani, White,
  Pr{\v{s}}a, {\v{Z}}ivkovi{\'c}, Berger, Magrez, Liu, Han, and
  R{\o}nnow}}]{Omrani2014}
\bibinfo{author}{\bibfnamefont{A.}~\bibnamefont{Omrani}},
  \bibinfo{author}{\bibfnamefont{J.}~\bibnamefont{White}},
  \bibinfo{author}{\bibfnamefont{K.}~\bibnamefont{Pr{\v{s}}a}},
  \bibinfo{author}{\bibfnamefont{I.}~\bibnamefont{{\v{Z}}ivkovi{\'c}}},
  \bibinfo{author}{\bibfnamefont{H.}~\bibnamefont{Berger}},
  \bibinfo{author}{\bibfnamefont{A.}~\bibnamefont{Magrez}},
  \bibinfo{author}{\bibfnamefont{Y.-H.} \bibnamefont{Liu}},
  \bibinfo{author}{\bibfnamefont{J.}~\bibnamefont{Han}}, \bibnamefont{and}
  \bibinfo{author}{\bibfnamefont{H.~M.} \bibnamefont{R{\o}nnow}},
  \bibinfo{journal}{Phys. Rev. B} \textbf{\bibinfo{volume}{89}},
  \bibinfo{pages}{064406} (\bibinfo{year}{2014}).

\bibitem[{\citenamefont{White et~al.}(2014)\citenamefont{White, Pr{\v{s}}a,
  Huang, Omrani, {\v{Z}}ivkovi{\'c}, Bartkowiak, Berger, Magrez, Gavilano, Nagy
  et~al.}}]{White2014}
\bibinfo{author}{\bibfnamefont{J.}~\bibnamefont{White}},
  \bibinfo{author}{\bibfnamefont{K.}~\bibnamefont{Pr{\v{s}}a}},
  \bibinfo{author}{\bibfnamefont{P.}~\bibnamefont{Huang}},
  \bibinfo{author}{\bibfnamefont{A.}~\bibnamefont{Omrani}},
  \bibinfo{author}{\bibfnamefont{I.}~\bibnamefont{{\v{Z}}ivkovi{\'c}}},
  \bibinfo{author}{\bibfnamefont{M.}~\bibnamefont{Bartkowiak}},
  \bibinfo{author}{\bibfnamefont{H.}~\bibnamefont{Berger}},
  \bibinfo{author}{\bibfnamefont{A.}~\bibnamefont{Magrez}},
  \bibinfo{author}{\bibfnamefont{J.}~\bibnamefont{Gavilano}},
  \bibinfo{author}{\bibfnamefont{G.}~\bibnamefont{Nagy}}, \bibnamefont{et~al.},
  \bibinfo{journal}{Phys. Rev. Lett.} \textbf{\bibinfo{volume}{113}},
  \bibinfo{pages}{107203} (\bibinfo{year}{2014}).

\bibitem[{\citenamefont{White et~al.}(2012)\citenamefont{White, Levati{\'c},
  Omrani, Egetenmeyer, Pr{\v{s}}a, {\v{Z}}ivkovi{\'c}, Gavilano, Kohlbrecher,
  Bartkowiak, Berger et~al.}}]{White2012}
\bibinfo{author}{\bibfnamefont{J.~S.} \bibnamefont{White}},
  \bibinfo{author}{\bibfnamefont{I.}~\bibnamefont{Levati{\'c}}},
  \bibinfo{author}{\bibfnamefont{A.}~\bibnamefont{Omrani}},
  \bibinfo{author}{\bibfnamefont{N.}~\bibnamefont{Egetenmeyer}},
  \bibinfo{author}{\bibfnamefont{K.}~\bibnamefont{Pr{\v{s}}a}},
  \bibinfo{author}{\bibfnamefont{I.}~\bibnamefont{{\v{Z}}ivkovi{\'c}}},
  \bibinfo{author}{\bibfnamefont{J.}~\bibnamefont{Gavilano}},
  \bibinfo{author}{\bibfnamefont{J.}~\bibnamefont{Kohlbrecher}},
  \bibinfo{author}{\bibfnamefont{M.}~\bibnamefont{Bartkowiak}},
  \bibinfo{author}{\bibfnamefont{H.}~\bibnamefont{Berger}},
  \bibnamefont{et~al.}, \bibinfo{journal}{J. Phys.: Cond. Matt.}
  \textbf{\bibinfo{volume}{24}}, \bibinfo{pages}{432201}
  (\bibinfo{year}{2012}).

\bibitem[{foo()}]{footnote1}
\bibinfo{note}{Splitting of bound states from continuous bands is accomplished
  by a redistribution of density of states within the bands, which is of
  importance here.}

\bibitem[{\citenamefont{Cooper et~al.}(1995)\citenamefont{Cooper, Khare, and
  Sukhatme}}]{Cooper1995}
\bibinfo{author}{\bibfnamefont{F.}~\bibnamefont{Cooper}},
  \bibinfo{author}{\bibfnamefont{A.}~\bibnamefont{Khare}}, \bibnamefont{and}
  \bibinfo{author}{\bibfnamefont{U.}~\bibnamefont{Sukhatme}},
  \bibinfo{journal}{Physics Reports} \textbf{\bibinfo{volume}{251}},
  \bibinfo{pages}{267} (\bibinfo{year}{1995}).

\bibitem[{\citenamefont{Klauder}(1979)}]{Klauder1979}
\bibinfo{author}{\bibfnamefont{J.~R.} \bibnamefont{Klauder}},
  \bibinfo{journal}{Phys. Rev. D} \textbf{\bibinfo{volume}{19}},
  \bibinfo{pages}{2349} (\bibinfo{year}{1979}).

\bibitem[{\citenamefont{Stone}(1989)}]{Stone1989}
\bibinfo{author}{\bibfnamefont{M.}~\bibnamefont{Stone}},
  \bibinfo{journal}{Nucl. Phys. B} \textbf{\bibinfo{volume}{314}},
  \bibinfo{pages}{557} (\bibinfo{year}{1989}).

\bibitem[{\citenamefont{Stone}(1996)}]{Stone1996}
\bibinfo{author}{\bibfnamefont{M.}~\bibnamefont{Stone}},
  \bibinfo{journal}{Phys. Rev. B} \textbf{\bibinfo{volume}{53}},
  \bibinfo{pages}{16573} (\bibinfo{year}{1996}).

\bibitem[{\citenamefont{Schulz et~al.}(2012)\citenamefont{Schulz, Ritz, Bauer,
  Halder, Wagner, Franz, Pfleiderer, Everschor, Garst, and Rosch}}]{Schulz2012}
\bibinfo{author}{\bibfnamefont{T.}~\bibnamefont{Schulz}},
  \bibinfo{author}{\bibfnamefont{R.}~\bibnamefont{Ritz}},
  \bibinfo{author}{\bibfnamefont{A.}~\bibnamefont{Bauer}},
  \bibinfo{author}{\bibfnamefont{M.}~\bibnamefont{Halder}},
  \bibinfo{author}{\bibfnamefont{M.}~\bibnamefont{Wagner}},
  \bibinfo{author}{\bibfnamefont{C.}~\bibnamefont{Franz}},
  \bibinfo{author}{\bibfnamefont{C.}~\bibnamefont{Pfleiderer}},
  \bibinfo{author}{\bibfnamefont{K.}~\bibnamefont{Everschor}},
  \bibinfo{author}{\bibfnamefont{M.}~\bibnamefont{Garst}}, \bibnamefont{and}
  \bibinfo{author}{\bibfnamefont{A.}~\bibnamefont{Rosch}},
  \bibinfo{journal}{Nat. Phys.} \textbf{\bibinfo{volume}{8}},
  \bibinfo{pages}{301} (\bibinfo{year}{2012}).

\bibitem[{\citenamefont{Everschor et~al.}(2011)\citenamefont{Everschor, Garst,
  Duine, and Rosch}}]{Everschor2011}
\bibinfo{author}{\bibfnamefont{K.}~\bibnamefont{Everschor}},
  \bibinfo{author}{\bibfnamefont{M.}~\bibnamefont{Garst}},
  \bibinfo{author}{\bibfnamefont{R.}~\bibnamefont{Duine}}, \bibnamefont{and}
  \bibinfo{author}{\bibfnamefont{A.}~\bibnamefont{Rosch}},
  \bibinfo{journal}{Phys. Rev. B} \textbf{\bibinfo{volume}{84}},
  \bibinfo{pages}{064401} (\bibinfo{year}{2011}).

\bibitem[{\citenamefont{Everschor et~al.}(2012)\citenamefont{Everschor, Garst,
  Binz, Jonietz, M{\"u}hlbauer, Pfleiderer, and Rosch}}]{Everschor2012}
\bibinfo{author}{\bibfnamefont{K.}~\bibnamefont{Everschor}},
  \bibinfo{author}{\bibfnamefont{M.}~\bibnamefont{Garst}},
  \bibinfo{author}{\bibfnamefont{B.}~\bibnamefont{Binz}},
  \bibinfo{author}{\bibfnamefont{F.}~\bibnamefont{Jonietz}},
  \bibinfo{author}{\bibfnamefont{S.}~\bibnamefont{M{\"u}hlbauer}},
  \bibinfo{author}{\bibfnamefont{C.}~\bibnamefont{Pfleiderer}},
  \bibnamefont{and} \bibinfo{author}{\bibfnamefont{A.}~\bibnamefont{Rosch}},
  \bibinfo{journal}{Phys. Rev. B} \textbf{\bibinfo{volume}{86}},
  \bibinfo{pages}{054432} (\bibinfo{year}{2012}).

\bibitem[{\citenamefont{Seki et~al.}(2012{\natexlab{b}})\citenamefont{Seki,
  Ishiwata, and Tokura}}]{Seki2012_2}
\bibinfo{author}{\bibfnamefont{S.}~\bibnamefont{Seki}},
  \bibinfo{author}{\bibfnamefont{S.}~\bibnamefont{Ishiwata}}, \bibnamefont{and}
  \bibinfo{author}{\bibfnamefont{Y.}~\bibnamefont{Tokura}},
  \bibinfo{journal}{Phys. Rev. B} \textbf{\bibinfo{volume}{86}},
  \bibinfo{pages}{060403} (\bibinfo{year}{2012}{\natexlab{b}}).

\bibitem[{\citenamefont{Heinze et~al.}(2011)\citenamefont{Heinze, von Bergmann,
  Menzel, Brede, Kubetzka, Wiesendanger, Bihlmayer, and
  Bl{\"u}gel}}]{heinze_spontaneous_2011}
\bibinfo{author}{\bibfnamefont{S.}~\bibnamefont{Heinze}},
  \bibinfo{author}{\bibfnamefont{K.}~\bibnamefont{von Bergmann}},
  \bibinfo{author}{\bibfnamefont{M.}~\bibnamefont{Menzel}},
  \bibinfo{author}{\bibfnamefont{J.}~\bibnamefont{Brede}},
  \bibinfo{author}{\bibfnamefont{A.}~\bibnamefont{Kubetzka}},
  \bibinfo{author}{\bibfnamefont{R.}~\bibnamefont{Wiesendanger}},
  \bibinfo{author}{\bibfnamefont{G.}~\bibnamefont{Bihlmayer}},
  \bibnamefont{and}
  \bibinfo{author}{\bibfnamefont{S.}~\bibnamefont{Bl{\"u}gel}},
  \bibinfo{journal}{Nat. Phys.} \textbf{\bibinfo{volume}{7}},
  \bibinfo{pages}{713} (\bibinfo{year}{2011}).

\bibitem[{\citenamefont{Baum and Stern}(2012)}]{Baum2012}
\bibinfo{author}{\bibfnamefont{Y.}~\bibnamefont{Baum}} \bibnamefont{and}
  \bibinfo{author}{\bibfnamefont{A.}~\bibnamefont{Stern}},
  \bibinfo{journal}{Phys. Rev. B} \textbf{\bibinfo{volume}{86}},
  \bibinfo{pages}{195116} (\bibinfo{year}{2012}).

\end{thebibliography}

\end{document}